\documentclass[twocolumn,american,prb, aps,final,superscriptaddress]{revtex4-1}
\usepackage[T1]{fontenc}
\usepackage[latin9]{inputenc}
\usepackage{xcolor}
\usepackage{pdfcolmk}
\usepackage{babel}
\usepackage{endnotes}
\usepackage{amsmath}
\usepackage{stackrel}
\usepackage{graphicx}
\PassOptionsToPackage{normalem}{ulem}
\usepackage{ulem}
\usepackage[unicode=true,
 bookmarks=true,bookmarksnumbered=false,bookmarksopen=false,
 breaklinks=false,pdfborder={0 0 1},backref=false,colorlinks=false]
 {hyperref}
\hypersetup{pdftitle={Ewald summation for ferroelectric perovksites with charges and dipoles},
 pdfauthor={Dawei Wang},
 pdfsubject={Physics},
 pdfkeywords={ferroelectric, perovskites,Ewald method}}

\makeatletter

\providecommand{\tabularnewline}{\\}
\providecolor{lyxadded}{rgb}{0,0,1}
\providecolor{lyxdeleted}{rgb}{1,0,0}

\DeclareRobustCommand{\lyxsout}[1]{\ifx\\#1\else\sout{#1}\fi}

\let\footnote=\endnote

\usepackage{times}
\allowdisplaybreaks
\usepackage{soul}
\usepackage{xcolor}

\makeatother

\begin{document}
\title{Ewald summation for ferroelectric perovskites with charges and dipoles }
\author{D. Wang}
\email{dawei.wang@xjtu.edu.cn}

\affiliation{School of Microelectronics \& State Key Laboratory for Mechanical
Behavior of Materials, Xi\textquoteright an Jiaotong University, Xi\textquoteright an
710049, China}
\author{J. Liu}
\affiliation{State Key Laboratory for Mechanical Behavior of Materials, School
of Materials Science and Engineering, Xi\textquoteright an Jiaotong
University, Xi'an 710049, China}
\author{J. Zhang}
\affiliation{School of Microelectronics \& State Key Laboratory for Mechanical
Behavior of Materials, Xi\textquoteright an Jiaotong University, Xi\textquoteright an
710049, China}
\author{S. Raza}
\affiliation{School of Microelectronics \& State Key Laboratory for Mechanical
Behavior of Materials, Xi\textquoteright an Jiaotong University, Xi\textquoteright an
710049, China}
\author{X. Chen}
\affiliation{Department of Applied Physics, Aalto University, Espoo 00076, Finland}
\affiliation{BroadBit Batteries Oy, Espoo 02150, Finland}
\author{C.-L. Jia}
\affiliation{School of Microelectronics \& State Key Laboratory for Mechanical
Behavior of Materials, Xi\textquoteright an Jiaotong University, Xi\textquoteright an
710049, China}
\affiliation{\textsuperscript{}Peter Gr�nberg Institute and Ernst Ruska Center
for Microscopy and Spectroscopy with Electrons, Research Center J�lich,
D-52425 J�lich, Germany}
\date{\today}
\begin{abstract}
\textcolor{black}{Ewald summation is an important technique used to
deal with long-range Coulomb interaction. While it is widely used
in simulations of molecules and solid state materials, many important
results are dispersed in literature and their implementations are
often buried deep in large software packages. Since reliable and systematic
calculation of Coulomb interaction is critical for the investigation
of perovskites, here we start from the fundamentals of Ewald summation
and derive clear expressions for long-range charge-charge, dipole-dipole,
and charge-dipole interactions, which can be readily used for numerical
computations. We also provide the interaction matrix for efficient
Monte Carlo simulations involving charges and dipoles, implementing
them in a Python software package. A new type of interaction matrix,
which accounts for the electrostatic energy change when ions are displaced,
is also derived and implemented. These results are the foundations
for the investigation of ferroelectric perovskites. }
\end{abstract}
\maketitle

\section{Introduction}

In the investigation of perovksites, long-range Coulomb interactions
are important since they are often the driving force of spontaneous
polarization and ferroelectricity \citep{Cohen1992,Zhong1995}. Accurate
and fast calculations of the dipole-dipole interactions are crucial
in such systems \citep{Zhong1995}. In addition to the dipole-dipole
interactions, complex perovksites (e.g., doping, alloying,defects,
and oxygen vacancy) \citep{Hageman1981,Xia2007,Hu2013,Tan2014,Bourguiba2016}
can introduce effective charges and their long-range interactions
in perovksites. Therefore, it also becomes inevitable to deal with
charge-charge and charge-dipole interactions. 

The difficulty to calculate the electrostatic energy in a system with
charges and dipoles lies in the long-range nature of Coulomb interaction.
The electrostatic energy due to the charge-charge interaction is given
by
\begin{align}
U_{\textrm{ES}}= & \frac{1}{2}\cdot\frac{1}{4\pi\varepsilon_{0}}\sum_{i,j}\frac{q_{i}q_{j}}{\left|\boldsymbol{r}_{i}-\boldsymbol{r}_{j}\right|},\label{eq:general-sum}
\end{align}
which decreases slowly with the distance between charges ($\sim1/r$),
making the convergence in numerical computations hard to achieve.
In the above expression, $\boldsymbol{r}_{i}$ is the position of
the charge $q_{i}$ and $\varepsilon_{0}$ is the vacuum permittivity.
In fact, the above series is conditionally convergent, i.e., the result
of the sum depends on the order of terms \citealp{Borwein1985,Min2012}.
The justification for using the Ewald method to treat this sum is
discussed in Refs. \onlinecite{Campbell1963,Borwein1985}.

The periodic boundary condition is usually adopted to enable the use
of the Ewald method. For infinite and periodic systems, the above
expression can be changed by organizing charges into \emph{supercells}

\begin{equation}
U_{\textrm{ES}}=\frac{1}{2}\left(\frac{1}{4\pi\varepsilon_{0}}\right)\sum_{\boldsymbol{R_{n}}}'\left(\sum_{i=1}^{N}\sum_{j=1}^{N}\frac{q_{i}q_{j}}{\left|\boldsymbol{r}_{ij}-\boldsymbol{R_{n}}\right|}\right)\label{eq:supercell-sum}
\end{equation}
where $\boldsymbol{r}_{ij}\equiv\boldsymbol{r}_{i}-\boldsymbol{r}_{j}$,
$\boldsymbol{n}=\left(n_{x},n_{y},n_{z}\right)$ with $n_{x}$, $n_{y}$,$n_{z}$
being integers. The summation over $i,j$ is within one supercell
containing $N$ charges, and $\boldsymbol{R}_{\boldsymbol{n}}$ denotes
the shift vector to other supercells. By running over all $\boldsymbol{R_{n}}$,
the summation covers all charges in the system by repeating the supercell.
The notation $\sum_{\boldsymbol{R_{n}}}'$ signifies that the term
where $i=j$ is omitted when $\boldsymbol{R_{n}}=0$. The charge at
$\boldsymbol{r}_{j}+\boldsymbol{R_{n}}$ is often called an image
charge of the charge at $\boldsymbol{r}_{j}$ in the first supercell.
In later sections, expressions similar to Eq. (\ref{eq:supercell-sum})
will be used.

Straightforward computation of Eq. (\ref{eq:general-sum}) or Eq.
(\ref{eq:supercell-sum}) is expensive due to the slow convergence.
One solution, which is also the first one of its type, is known as
the \textit{\textcolor{black}{Ewald method}} introduced \textcolor{black}{in
1921 by Paul P. Ewald \citep{Ewald1921}. The crucial insight in this
approach is to split the sum into two parts, which are treated differently
(sum in real and reciprocal spaces) to achieve fast convergence. More
specifically, it is necessary to choose a function $f\left(r\right)$
with $r\equiv\left|\boldsymbol{r}\right|$ so that }

\begin{equation}
\frac{1}{r}=\frac{f\left(r\right)}{r}+\frac{1-f\left(r\right)}{r},\label{eq:f-r}
\end{equation}
where $f\left(r\right)/r$ shall decay faster with $r$ than $1/r$
(to sum efficiently in real space), while $\left[1-f\left(r\right)\right]/r$
decays slow in real space (sum efficiently in the reciprocal space).
As we will see below, one choice for $f\left(r\right)$ is the \emph{complementary
error function}.

The Ewald method is a special type of Poisson solver \citep{Possion_solver}
as it provides the electrostatic potential for a given charge distribution.
The Ewald summation is also an application of the more general \emph{Poisson
summation} \citep{Stein2003}. The formula shown in Eq. (\ref{eq:P-J_relation})
is an example of the Poisson summation formula, which is also used
in Ref. \onlinecite{Grzybowski2000} for systems having periodicity
in less than three dimensions.

In an Ewald summation, the reciprocal space is the critical part.
In particular, the calculation of terms like $U\left(\boldsymbol{k}\right)$
in Eq (\ref{eq:U_k}), which represents the charge/dipole distribution
in the reciprocal space, needs to be efficiently calculated. In the
past, people have tried to optimize the calculation of them, giving
rise to the Particle--Particle-Particle--Mesh (P$^{3}$M) and Particle-Mesh-Ewald
(PME) methods for arbitrary distribution of charges. The main idea
is to exploit the Fast Fourier Transformation (FFT) in the evaluation
terms such as Eq. (\ref{eq:U_k}). However, given the arbitrary positions
of $q_{i}$ (within a given supercell), an interpolation of the charges
to a regular mesh is necessary, which results in the PME method \citealp{Darden1993}.
Another popular method, proposed before PME and also strongly depending
on the efficiency of FFT, is the P$^{3}$M method \citealp{Hockney1988}
which use both the direct sum of particle-particle interaction (for
particles close to each other) and the particle-mesh method (to treat
particles separated far away). Since provskites already provide us
with a regular mesh (thanks to the Bravais lattice), we do not need
to use PME or P$^{3}$M, but will focus on the interaction matrix
{[}e.g., Eq. (\ref{eq:charge-charge_matrix}){]}, which is necessary
for efficient Monte-Carlo (MC) and Molecular Dynamics (MD) simulations
of perovskites \citep{Zhong1995,Akbarzadeh2012,Jiang2015,Wang2016}.

The aim of this work is twofold: (i) Summarize essential results involving
Ewald summation dispersed in literature, providing detailed and reliable
expressions necessary for numerical implementation. We will cover
charge-charge, dipole-dipole, and charge-dipole interactions, needed
for understanding complex perovskites with alloying, doping, and defects.
Moreover, we also derive the electrostatic energy expression when
ions have small displacements (which can cause polarization); (ii)
Discuss and provide numerical implementations for each type of interactions
mentioned in (i). In particular, we provide Python \citep{Python}
programs that generate the interaction matrix in the \texttt{netcdf}
format \citep{netcdf} and consider other less investigated interactions,
including the charge-dipole interaction and interactions due to small
charge displacements. 

In this work, we deal with bulk materials having periodicity in all
three dimensions. Ewald summation for finite extent in two or three
dimensions is discussed in Refs. {[}\onlinecite{Grzybowski2000,Porto2000,Santos2016}{]}.
Such results are necessary for the investigation of nanowires and
thin films. It is also worth noting that the Fourier transformation
of the dipole interaction matrix can also be accelerated using the
Ewald method as discussed in Ref. {[}\onlinecite{Bowden1981}{]}. 

This paper is organized as the follows. In Sec. \ref{sec:charge-charge},
we treat charge-charge interactions, providing details to explain
the Ewald method and preparing for other type of interactions. In
Sec. \ref{sec:Other-interaction-matrices}, we obtain the dipole-dipole
and dipole-charge interaction matrices. In addition, we also consider
how Coulomb energy changes when ions are displaced. In Sec. \ref{sec:Implementation},
we provide information regarding the implementation using Python.
In Sec. \ref{sec:Discussion}, we extend the Ewald summation to general
Bravais lattices, discuss several less important issues, and provide
more details of derivations used in previous sections. Finally in
Sec. \ref{sec:Conclusion}, we give a brief summary.

\section{Charge-charge interaction \label{sec:charge-charge}}

We first discuss the charge-charge interaction energy with some details
as the results obtained here can serve as starting points for dipole-dipole
and charge-dipole interactions. For simplicity, hereafter simple cubic
lattice is used with the length of the supercell to be \textit{$L$}.
General Bravais lattice will be discussed in Sec. \ref{sec:General-Bravai-lattice},
which shows a natural transition from the simple cubic lattice is
possible. Throughout this paper, we use Latin letters (e.g., $i$,
$j$) to index charges or dipoles, and Greek letters (e.g., $\alpha$,$\beta$)
to indicate Cartesian directions ($x,y,z$). We also use $V$ to denote
the volume of the whole crystal and $\Omega$ to indicate the volume
of the chosen supercell.

\subsection{Separated potentials}

A key step in the Ewald method is to split the Coulomb interaction
into long-range and short-range terms. We first focus on the charge
distribution of a point charge $q_{i}$ at $\boldsymbol{r}=0$ with
the charge distribution 
\begin{align*}
\rho_{i}\left(\boldsymbol{r}\right)= & \delta\left(\boldsymbol{r}\right),
\end{align*}
which generates a Coulomb potential that slowly decays with $1/r$.
Here $\delta\left(\boldsymbol{r}\right)$ is a Dirac-delta function.
The charge distribution can be split into two terms in the following
way
\begin{align*}
\rho_{i}\left(\boldsymbol{r}\right)= & \rho_{i}^{S}\left(\boldsymbol{r}\right)+\rho_{i}^{L}\left(\boldsymbol{r}\right),
\end{align*}
where
\begin{align*}
\rho_{i}^{S}\left(\boldsymbol{r}\right) & =q_{i}\delta\left(\boldsymbol{r}\right)-q_{i}G_{\sigma}\left(\boldsymbol{r}\right),\\
\rho_{i}^{L}\left(\boldsymbol{r}\right) & =q_{i}G_{\sigma}\left(\boldsymbol{r}\right),
\end{align*}
and 

\[
G_{\sigma}\left(\boldsymbol{r}\right)=\frac{1}{\left(2\pi\sigma^{2}\right)^{3/2}}\textrm{exp}\left(-\frac{\left|\boldsymbol{r}\right|^{2}}{2\sigma^{2}}\right),
\]
with $\sigma$ being a constant specifying the spread of the Gaussian
function. The physical meaning here is to intentionally introduce
a smooth charge distribution that neutralize the point charge at $\boldsymbol{r}=0$,
resulting in $\rho_{i}^{S}\left(\boldsymbol{r}\right)$, which only
produces short-range electric potentials (which can be dealt with
in real space), and an additional charge distribution $\rho_{i}^{L}\left(\boldsymbol{r}\right)$
(which can be dealt with in the reciprocal space).

The potential field generated by $G_{\sigma}\left(\boldsymbol{r}\right)$
can be obtained by solving the Poisson equation,

\begin{equation}
\nabla^{2}\phi_{\sigma}\left(\boldsymbol{r}\right)=-\frac{G_{\sigma}\left(\boldsymbol{r}\right)}{\varepsilon_{0}},\label{eq:possion-cartesian}
\end{equation}
where the spherical symmetry of $G_{\sigma}\left(\boldsymbol{r}\right)$
is used to have
\begin{align*}
\frac{1}{r}\frac{\partial^{2}}{\partial r^{2}}\left(r\phi_{\sigma}\left(r\right)\right)= & -\frac{G_{\sigma}\left(\boldsymbol{r}\right)}{\varepsilon_{0}}.
\end{align*}
Integration over $r$ gives 

\begin{equation}
\frac{\partial}{\partial r}\left(r\phi_{\sigma}\left(r\right)\right)=-\int_{r}^{\infty}\frac{G_{\sigma}\left(r^{\prime}\right)}{\varepsilon_{0}}r^{\prime}dr^{\prime}=-\frac{\sigma^{2}}{\varepsilon_{0}}G_{\sigma}\left(r\right),\label{eq:possion-spherical}
\end{equation}
which results in 
\[
\phi_{\sigma}\left(\boldsymbol{r}\right)=\frac{1}{4\pi\varepsilon_{0}r}\textrm{erf}\left(\frac{r}{\sqrt{2}\sigma}\right),
\]
where $\textrm{erf}\left(x\right)=\frac{2}{\sqrt{\pi}}\int_{0}^{x}e^{-\eta^{2}}d\eta$.

\begin{figure}
\begin{centering}
\includegraphics[width=7cm]{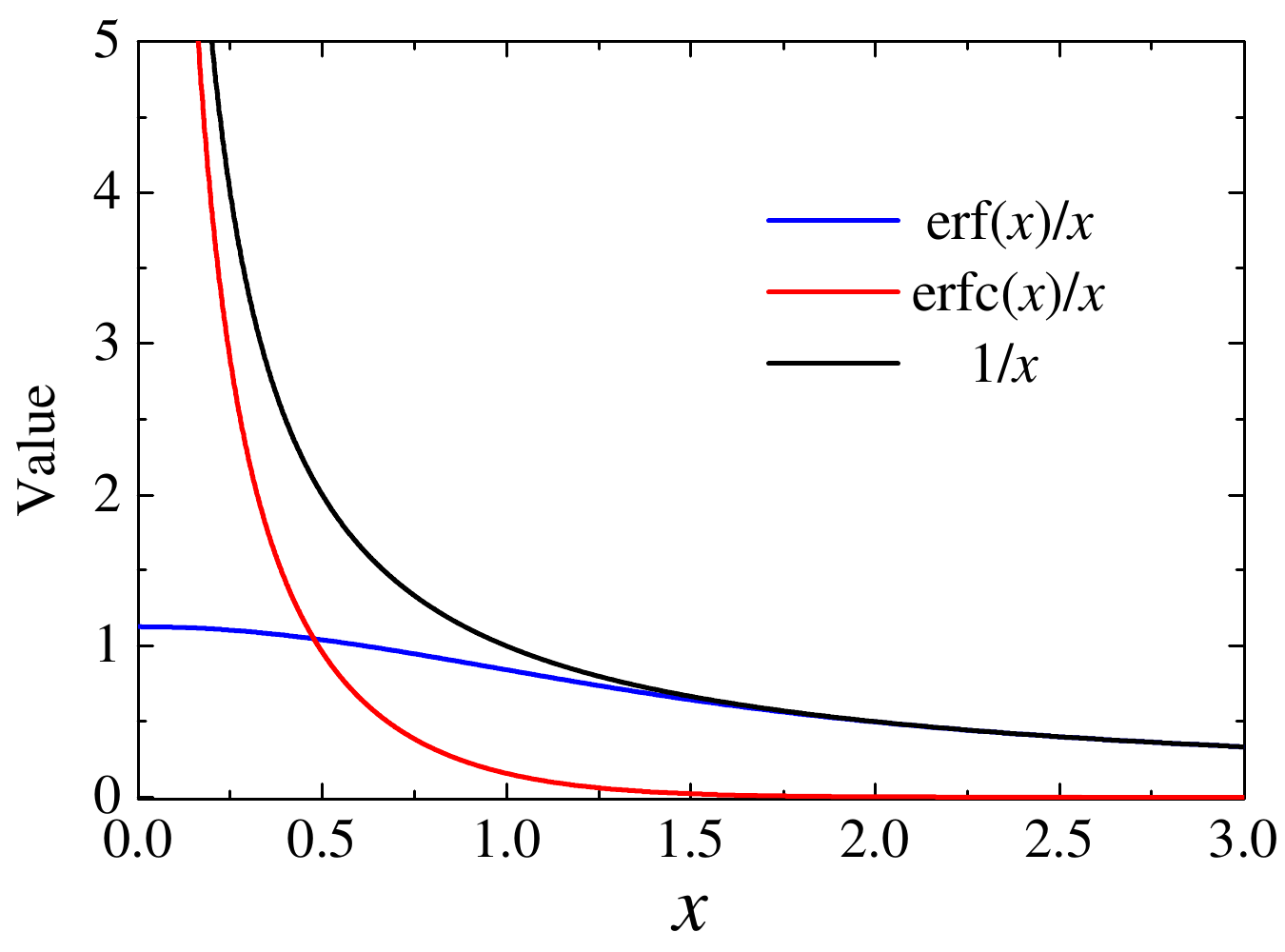}
\par\end{centering}
\caption{Comparison of the $\textrm{erf}\left(x\right)/x$, $\textrm{erfc}\left(x\right)/x$,
and $1/x$ functions. \label{fig:The-graph-of-erf}}
\end{figure}
Therefore, the electrostatic potentials generated by $\rho_{i}^{L}\left(\boldsymbol{r}\right)$
and $\rho_{i}^{S}\left(\boldsymbol{r}\right)$ are 

\begin{align}
\phi_{i}^{L}\left(\boldsymbol{r}\right)= & \frac{1}{4\pi\varepsilon_{0}}\frac{q_{i}}{\left|\boldsymbol{r}-\boldsymbol{r}_{i}\right|}\textrm{erf}\left(\frac{\left|\boldsymbol{r}-\boldsymbol{r}_{i}\right|}{\sqrt{2}\sigma}\right),\label{eq:Reciprical space}\\
\phi_{i}^{S}\left(\boldsymbol{r}\right)= & \frac{1}{4\pi\varepsilon_{0}}\frac{q_{i}}{\left|\boldsymbol{r}-\boldsymbol{r}_{i}\right|}\textrm{erfc}\left(\frac{\left|\boldsymbol{r}-\boldsymbol{r}_{i}\right|}{\sqrt{2}\sigma}\right),\label{eq:Real space}
\end{align}
where $\textrm{erf}\left(x\right)$ is the error function and $\textrm{erfc}\left(x\right)=1-\textrm{erf}\left(x\right)$
is the complementary error function. The three functions $\textrm{erf\ensuremath{\left(x\right)}}/x$
, $\textrm{erfc\ensuremath{\left(x\right)}}/x$ and $1/x$ are plotted
in Fig. \ref{fig:The-graph-of-erf}, which indicates that $\text{erfc}\left(x\right)/x$
decays faster than $1/x$ and $\textrm{erf}\left(x\right)/x$ decays
similar to $1/x$ for large $x$.

\subsection{$E^{S}$ and $E^{L}$\label{subsec:Es-and-El}}

Given the potentials $\phi_{i}^{S}\left(\boldsymbol{r}\right)$ and
$\phi_{i}^{L}\left(\boldsymbol{r}\right)$, the energy can also be
split into two parts. Similar to the expression in Eq. (\ref{eq:supercell-sum}),
the short-range potential due to all charges is given by
\begin{align*}
\Phi^{S}\left(\boldsymbol{r}\right) & =\frac{1}{4\pi\varepsilon_{0}}\underset{\boldsymbol{n}}{\sum}\stackrel[j=1]{N}{\sum}q_{j}\phi_{j}\left(\boldsymbol{r}-\boldsymbol{r}_{j}-\boldsymbol{n}L\right)\\
 & =\frac{1}{4\pi\varepsilon_{0}}\underset{\boldsymbol{n}}{\sum}\stackrel[j=1]{N}{\sum}\frac{q_{j}}{\left|\boldsymbol{r}-\boldsymbol{r}_{j}-\boldsymbol{n}L\right|}\textrm{erfc}\left(\frac{\left|\boldsymbol{r}-\boldsymbol{r}_{j}-\boldsymbol{n}L\right|}{\sqrt{2}\sigma}\right).
\end{align*}
Therefore the electrostatic energy $E^{S}$ of the system is given
by

\begin{eqnarray}
E^{S} & = & \frac{1}{2}\stackrel[i=1]{N}{\sum}q_{i}\Phi^{S}\left(\boldsymbol{r}_{i}\right)\nonumber \\
 & = & \frac{1}{2}\left(\frac{1}{4\pi\varepsilon_{0}}\right)\sum_{\boldsymbol{n}}'\stackrel[i=1]{N}{\sum}\stackrel[j=1]{N}{\sum}\frac{q_{i}q_{j}}{\left|\boldsymbol{r}_{i}-\boldsymbol{r}_{j}-\boldsymbol{n}L\right|}\nonumber \\
 &  & \times\textrm{erfc}\left(\frac{\left|\boldsymbol{r}_{i}-\boldsymbol{r}_{j}-\boldsymbol{n}L\right|}{\sqrt{2}\sigma}\right),\label{eq:E_S}
\end{eqnarray}
where the overall $1/2$ is necessary to avoid double counting. We
note the above expression is the same as Eq. (\ref{eq:supercell-sum})
except the extra factor of $\textrm{erfc}\left(\left|\boldsymbol{r}_{i}-\boldsymbol{r}_{j}-\boldsymbol{n}L\right|/\sqrt{2}\sigma\right)$.
Because $\textrm{erfc}\left(x\right)/x$ decays faster, $E^{S}$ can
be efficiently calculated in real space. In other words, Eq. (\ref{eq:E_S})
can be directly used in numerical implementation without further transformation.
We note that Eq. (\ref{eq:E_S}) represents the electrostatic energy
due to charge-charge interaction of a supercell.

The electrostatic energy due to the long-range potential is given
by
\begin{eqnarray}
E^{L} & = & \frac{1}{2}\stackrel[i=1]{N}{\sum}q_{i}\Phi^{L}\left(\boldsymbol{r}_{i}\right)\nonumber \\
 & = & \frac{1}{2}\frac{1}{4\pi\varepsilon_{0}}\sum_{\boldsymbol{n}}'\stackrel[i=1]{N}{\sum}\stackrel[j=1]{N}{\sum}\frac{q_{i}q_{j}}{\left|\boldsymbol{r}_{i}-\boldsymbol{r}_{j}-\boldsymbol{n}L\right|}\nonumber \\
 &  & \times\textrm{erf}\left(\frac{\left|\boldsymbol{r}_{i}-\boldsymbol{r}_{j}-\boldsymbol{n}L\right|}{\sqrt{2}\sigma}\right).\label{eq:E_L}
\end{eqnarray}
To proceed further (see below), we can use the regular $\sum_{n}$ 

\begin{align}
E^{L}= & \frac{1}{2}\frac{1}{4\pi\varepsilon_{0}}\sum_{i=1}^{N}\sum_{j=1}^{N}q_{i}q_{j}\sum_{\boldsymbol{n}}\frac{q_{j}}{\left|\boldsymbol{r}_{i}-\boldsymbol{r}_{j}-\boldsymbol{n}L\right|}\nonumber \\
 & \times\textrm{erf}\left(\frac{\left|\boldsymbol{r}_{i}-\boldsymbol{r}_{j}-\boldsymbol{n}L\right|}{\sqrt{2}\sigma}\right)\nonumber \\
 & -\frac{1}{2}\frac{1}{4\pi\varepsilon_{0}}\sum_{i=1}^{N}\frac{1}{\sigma}\sqrt{\frac{2}{\pi}}q_{i}^{2}\label{eq:E_L-2}
\end{align}
where in the last step the relation $\lim_{x\rightarrow0}\left(\textrm{erf}\left(x\right)/x\right)=2/\sqrt{\pi}$
has been used.

Since $\textrm{erf}\left(x\right)/x$ decays slowly ($\sim1/x$ for
large $x$, see Fig. \ref{fig:The-graph-of-erf}), the expression
in Eq. (\ref{eq:E_L-2}) needs to be converted to a sum in the reciprocal
space for fast numerical convergence. The core term in Eq. (\ref{eq:E_L-2})
has the form
\begin{align}
f\left(\boldsymbol{r}\right)= & \sum_{\boldsymbol{n}}\frac{1}{\left|\boldsymbol{r}-\boldsymbol{n}L\right|}\textrm{erf}\left(\frac{\left|\boldsymbol{r}-\boldsymbol{n}L\right|}{\sqrt{2}\sigma}\right),\label{eq:f_r}
\end{align}
which is periodic as $f\left(\boldsymbol{r}+\boldsymbol{n}L\right)=f\left(\boldsymbol{r}\right)$.
Therefore it can be expanded into a Fourier series,
\begin{align*}
f\left(\boldsymbol{r}\right)= & \sum_{\boldsymbol{k}}a_{\boldsymbol{k}}\exp\left(i\boldsymbol{k}\cdot\boldsymbol{r}\right),
\end{align*}
where $\boldsymbol{k}=2\pi\left(m_{1},m_{2},m_{3}\right)/L$ with
$m_{1,2,3}$ being integers. We perform the inverse Fourier transformation
to have (see Sec. \ref{subsec:Derivation-of-alpha-k} for detailed
derivation),
\begin{align*}
a_{k}= & \frac{4\pi}{\Omega k^{2}}\exp\left(-\frac{k^{2}\sigma^{2}}{2}\right),
\end{align*}
where $\Omega=L^{3}$ is the volume of the supercell. Therefore, the
interaction energy can be represented in the reciprocal space,
\begin{align}
E^{L}= & \frac{1}{2}\frac{1}{4\pi\varepsilon_{0}\Omega}\stackrel[i=1]{N}{\sum}\stackrel[j=1]{N}{\sum}q_{i}q_{j}\sum_{\boldsymbol{k}\neq0}\frac{4\pi e^{-\sigma^{2}k^{2}/2}}{k^{2}}\cos\left[\boldsymbol{k}\cdot\left(\boldsymbol{r}_{i}-\boldsymbol{r}_{j}\right)\right]\nonumber \\
 & -\frac{1}{4\pi\varepsilon_{0}}\frac{1}{\sqrt{2\pi}\sigma}\sum_{i=1}^{N}q_{i}^{2}.\label{eq:E_L-Ewald}
\end{align}
In the above equation, we have ignored the $\boldsymbol{k}=0$ term
and the reason is explained in Sec. \ref{subsec:Derivation-of-alpha-k}.

Finally the total energy due to the charge-charge interaction is given
by

\[
E_{\textrm{chg\text{\_}chg}}=\stackrel[ij]{N}{\sum}Q_{ij}q_{i}q_{j},
\]
where

\begin{align}
 & Q_{ij}\nonumber \\
= & \frac{1}{2}\frac{1}{4\pi\varepsilon_{0}}\left\{ \frac{1}{\left|\boldsymbol{r}_{i}-\boldsymbol{r}_{j}\right|}\textrm{erfc}\left(\frac{\left|\boldsymbol{r}_{i}-\boldsymbol{r}_{j}\right|}{\sqrt{2}\sigma}\right)\left(1-\delta_{ij}\right)\right.\nonumber \\
 & +\underset{n\neq0}{\sum}\frac{1}{\left|\boldsymbol{r}_{i}-\boldsymbol{r}_{j}-\boldsymbol{n}L\right|}\textrm{erfc}\left(\frac{\left|\boldsymbol{r}_{i}-\boldsymbol{r}_{j}-\boldsymbol{n}L\right|}{\sqrt{2}\sigma}\right)\nonumber \\
 & \left.+\frac{4\pi}{\Omega}\sum_{\boldsymbol{k}\neq0}\frac{e^{-\sigma^{2}k^{2}/2}}{k^{2}}\cos\left[\boldsymbol{k}\cdot\left(\boldsymbol{r}_{i}-\boldsymbol{r}_{j}\right)\right]-\frac{1}{\sigma}\sqrt{\frac{2}{\pi}}\delta_{ij}\right\} .\label{eq:charge-charge_matrix}
\end{align}

\subsection{General result and $\sigma$}

The derivation shown in Sec. \ref{subsec:Es-and-El} results in a
more general expression 
\begin{align}
 & \sum_{n}\frac{1}{\left|\boldsymbol{r}_{1}-\left(\boldsymbol{r}_{2}+\boldsymbol{n}L\right)\right|}\nonumber \\
= & \left[\underset{n}{\sum}\frac{1}{\left|\boldsymbol{r}_{1}-\boldsymbol{r}_{2}-\boldsymbol{n}L\right|}\textrm{erfc}\left(\frac{\left|\boldsymbol{r}_{1}-\boldsymbol{r}_{2}-\boldsymbol{n}L\right|}{\sqrt{2}\sigma}\right)\right.\nonumber \\
 & +\frac{4\pi}{\Omega}\sum_{\boldsymbol{k}\neq0}\frac{e^{-\sigma^{2}k^{2}/2}}{k^{2}}\cos\left[\boldsymbol{k}\cdot\left(\boldsymbol{r}_{1}-\boldsymbol{r}_{2}\right)\right]\nonumber \\
 & \left.-\lim_{r_{i}\rightarrow r_{j}}\frac{1}{\left|\boldsymbol{r}_{i}-\boldsymbol{r}_{j}\right|}\textrm{erf}\left(\frac{\left|\boldsymbol{r}_{i}-\boldsymbol{r}_{j}\right|}{\sqrt{2}\sigma}\right)\right].\label{eq:sum-1/r}
\end{align}
which can be used in the following sections for the dipole-dipole
and dipole-charge interactions. This result is consistent with Ziman's
result \citealp{Ziman} and will be extended to general Bravais lattice
in Sec. \ref{sec:General-Bravai-lattice}, where another approach
to reach this result is shown.

In principle, the value of $\sigma$ in Eq. (\ref{eq:sum-1/r}) can
be chosen arbitrarily. In practice, it shall be chosen to reduce the
summation in either the real space or the reciprocal space. It was
pointed out that the value of $\alpha=1/\sqrt{2}\sigma$ can be chosen
as $\alpha\approx\sqrt{-\ln\delta}$ so that the error is on the order
of $\delta$ (e.g., we can set $\delta=10^{-12}$) \citep{Perram1988,Toukmaji1996}.
We have adopted this choice in our numerical implementations and will
show that the real-space summation can be ignored in Sec. \ref{subsec:Summation-in-real}
\citep{sigma}.

\section{Other interaction matrices\label{sec:Other-interaction-matrices}}

The Ewald summation of charge-charge interaction provides us with
a basis to derive similar expressions for other types of interactions,
such as dipole-dipole and charge-dipole interactions. In obtaining
these expressions, point dipoles are assumed.\textbf{ }Point dipoles
had been employed to numerically simulate perovskites since the introduction
of the effective Hamiltonian approach \citep{Zhong1995}, generating
many important results and insights. \citep{Wang2013,Jiang2014,Jiang2015,Wang2016,Liu2018}

\subsection{Dipole-dipole interaction\label{sec:Dipole-dipole-interaction}}

The dipole-dipole interaction energy in a supercell is given by
\begin{align*}
E_{\textrm{dip-dip}}= & \frac{1}{2}\frac{1}{4\pi\varepsilon_{0}}\sum_{i\in\textrm{supercell}}\sum_{\alpha}\left(\boldsymbol{u}_{i}\right)_{\alpha}\\
 & \times\sum_{j\neq i}\sum_{\beta}\left[\frac{\delta_{\alpha,\beta}-3\left(\hat{\boldsymbol{r}}_{ij}\right)_{\alpha}\left(\hat{\boldsymbol{r}}_{ij}\right)_{\beta}}{r_{ij}^{3}}\right]\left(\boldsymbol{u}_{j}\right)_{\beta},
\end{align*}
where the sum over $i$ is inside the supercell while the sum over
$j$ expands to the whole space. Following Eq. (\ref{eq:supercell-sum}),
the above sum can be further converted to
\begin{align*}
 & E_{\textrm{dip-dip}}\\
= & \frac{1}{2}\frac{1}{4\pi\varepsilon_{0}}\sum_{\boldsymbol{n}}^{\prime}\sum_{i,j}\sum_{\alpha,\beta}\left(\boldsymbol{u}_{i}\right)_{\alpha}\left(\boldsymbol{u}_{j}\right)_{\beta}\\
 & \left[\frac{\delta_{\alpha,\beta}}{\left|\boldsymbol{r}_{ij}-\boldsymbol{n}L\right|^{3}}-\frac{3\left(\boldsymbol{r}_{ij}-\boldsymbol{n}L\right)_{\alpha}\left(\boldsymbol{r}_{ij}-\boldsymbol{n}L\right)_{\beta}}{\left|\boldsymbol{r}_{ij}-\boldsymbol{n}L\right|^{5}}\right],
\end{align*}
where both $i$ and $j$ belong to the supercell now.

To proceed further, we need to focus on the sum over $\boldsymbol{n}$.
For simplicity we use $\boldsymbol{r}=\boldsymbol{r}_{i}-\boldsymbol{r}_{j}$
to handle the above expression. A little algebra shows that
\begin{align}
 & \sum_{n}\left[\frac{\delta_{\alpha,\beta}}{\left|\boldsymbol{r}-\boldsymbol{n}L\right|^{3}}-\frac{3\left(\boldsymbol{r}-\boldsymbol{n}L\right)_{\alpha}\left(\boldsymbol{r}-\boldsymbol{n}L\right)_{\beta}}{\left|\boldsymbol{r}-\boldsymbol{n}L\right|^{5}}\right]\nonumber \\
= & -\partial_{r_{\alpha}}\partial_{r_{\beta}}\sum_{n}\frac{1}{\left|\boldsymbol{r}-\boldsymbol{n}L\right|}.\label{eq:important-1}
\end{align}
Using the result from Eq. (\ref{eq:sum-1/r}) and applying $\partial_{r_{\alpha}}\partial_{r_{\beta}}$
to it, we have
\begin{align}
 & E_{\textrm{dip-dip}}\label{eq:supercell-dip-dip-1}\\
= & \frac{1}{2}\frac{1}{4\pi\varepsilon_{0}}\sum_{i,j\in\textrm{supercell}}\sum_{\alpha,\beta}\left(\boldsymbol{u}_{i}\right)_{\alpha}\left(\boldsymbol{u}_{j}\right)_{\beta}\nonumber \\
 & \left\{ \sum_{n}\left.\frac{\sqrt{2}}{4\sigma^{3}x^{3}}\left[-\delta_{\alpha\beta}B_{\alpha\beta}\left(x\right)+\frac{x_{\alpha}x_{\beta}}{x^{2}}C_{\alpha\beta}\left(x\right)\right]\right|_{x=\frac{\left|\boldsymbol{r}-\boldsymbol{n}L\right|}{\sqrt{2}\sigma}}\right.\nonumber \\
 & \left.+\frac{4\pi}{\Omega}\sum_{\boldsymbol{k}\neq0}\frac{e^{-\sigma^{2}k^{2}/2}}{k^{2}}k_{\alpha}k_{\beta}\cos\left(\boldsymbol{k}\cdot\boldsymbol{r}\right)-\sqrt{\frac{2}{\pi}}\frac{1}{3\sigma^{3}}\delta_{\alpha\beta}\delta_{ij}\right\} ,\nonumber 
\end{align}
where $x=\frac{\left|\boldsymbol{r}-\boldsymbol{n}L\right|}{\sqrt{2}\sigma}$,
$B\left(x\right)=\frac{2}{\sqrt{\pi}}x\exp\left(-x^{2}\right)+\textrm{erfc}(x)$,
and $C\left(x\right)=\frac{2}{\sqrt{\pi}}\left(1+\frac{2x^{2}}{3}\right)x\exp\left(-x^{2}\right)+\textrm{erfc}(x)$.
For the use in MC simulation, this equation can be rewritten as 
\[
E_{\textrm{dip-dip}}=\stackrel[ij,\alpha\beta]{N}{\sum}Q_{ij\alpha\beta}\left(\boldsymbol{u}_{i}\right)_{\alpha}\left(\boldsymbol{u}_{j}\right)_{\beta}
\]
where 
\begin{align}
Q_{ij\alpha\beta}= & \frac{1}{2\pi\epsilon_{0}}\left[\frac{\pi}{\Omega}\sum_{k\neq0}\frac{1}{k^{2}}e^{-\frac{\sigma^{2}k^{2}}{2}}\right.\nonumber \\
 & \left.\times\textrm{cos}\left(\boldsymbol{k}\cdot\boldsymbol{r}_{ij}\right)k_{\alpha}k_{\beta}-\frac{\alpha^{3}}{3\sqrt{\pi}}\delta_{\alpha\beta}\delta_{ij}\right],\label{eq:dipole-dipole-matrix}
\end{align}
same as previous results \citep{Zhong1995}. In the above expression,
the sum in the real space is ignored and the reason is explained in
Sec. \ref{subsec:Summation-in-real}. 

\subsection{Charge-dipole interaction\label{sec:Charge-dipole-interaction}In }

Charge-dipole interaction had been discussed in Ref. {[}\onlinecite{Beck2010}{]},
which has different emphasis than ours. Similar to previous sections,
we focus on deriving the interaction matrix. Without the symmetry
between charge-charge (or dipole-dipole) interactions, two parts of
energy need to be considered, i.e., the energy of charge under the
electric potential of dipoles and \emph{vice versa}. Moreover, to
make sure that charges and dipoles are not on exactly the same position,
we assume dipoles are on the lattice sites, i.e., $\boldsymbol{r}_{i}$,
while charges are shifted to $\boldsymbol{r}_{i}+1/2\boldsymbol{a}_{1}+1/2\boldsymbol{a}_{2}+1/2\boldsymbol{a}_{3}$
where $\boldsymbol{a}_{1,2,3}$ are the Bravais lattice. For simple
cubic lattice, it is $\boldsymbol{r}_{i}+\left(1/2,1/2,1/2\right)a$
where $a$ is the lattice constant of the unit cell (note $L$ is
the lattice constant of the supercell). For simplicity, we use $\boldsymbol{d}$
to denote this shift $\boldsymbol{d}=\left(1/2,1/2,1/2\right)a$. 

The electrostatic potential given by a dipole $\boldsymbol{u}$ is\citep{Jackson1999}, 

\begin{align}
\phi_{i}\left(\boldsymbol{r}\right)=\frac{1}{4\pi\varepsilon_{0}}\frac{\boldsymbol{u}_{i}\cdot\boldsymbol{r}}{r^{3}}= & -\frac{\boldsymbol{u}_{i}}{4\pi\varepsilon_{0}}\cdot\nabla\left(\frac{1}{\boldsymbol{r}}\right)\label{eq:dip_potential-1}
\end{align}
Therefore, the energy for the charges under dipole potential is given
by 

\begin{equation}
E_{\textrm{chg-dip}}=\sum_{i}\left[-\sum_{\boldsymbol{n}}\sum_{j}\frac{q_{i}\boldsymbol{u}_{j}}{4\pi\varepsilon_{0}}\cdot\nabla_{\boldsymbol{r}_{ij}}\frac{1}{\left|\boldsymbol{r}_{ij}+\boldsymbol{d}-\boldsymbol{n}L\right|}\right].
\end{equation}
On the other hand, the potential energy of a dipole in an electric
field is given by \citep{Jackson1999}
\begin{align*}
U= & -\boldsymbol{u}\cdot\boldsymbol{E}\\
= & \boldsymbol{u}\cdot\nabla\phi\left(\boldsymbol{r}\right).
\end{align*}
Therefore, given the electric field from charges, this energy is

\begin{align*}
E_{\textrm{dip-chg}}= & \sum_{j}\left[\sum_{\boldsymbol{n}}\sum_{i}\frac{\boldsymbol{u}_{j}q_{i}}{4\pi\varepsilon_{0}}\cdot\nabla_{\boldsymbol{r}_{ji}}\frac{1}{\left|\boldsymbol{r}_{ji}-\boldsymbol{d}-\boldsymbol{n}L\right|}\right]\\
= & -\sum_{i}\left[\sum_{\boldsymbol{n}}\sum_{j}\frac{\boldsymbol{u}_{j}q_{i}}{4\pi\varepsilon_{0}}\cdot\nabla_{\boldsymbol{r}_{ij}}\frac{1}{\left|\boldsymbol{r}_{ij}+\boldsymbol{d}-\boldsymbol{n}L\right|}\right].
\end{align*}
Hence the total electrostatic energy between charge and dipole is
given by

\begin{align*}
E_{\textrm{CD}}= & \frac{1}{2}\left(E_{\textrm{dip-chg}}+E_{\textrm{dip-chg}}\right)\\
= & -\sum_{\boldsymbol{n}}\sum_{ij}\frac{q_{i}\boldsymbol{u}_{j}}{4\pi\varepsilon_{0}}\cdot\nabla_{\boldsymbol{r}_{ij}}\left(\frac{1}{\left|\boldsymbol{r}_{ij}+\boldsymbol{d}-\boldsymbol{n}L\right|}\right)
\end{align*}

Following Eq. \ref{eq:sum-1/r} , we obtain

\begin{align*}
 & \nabla_{\boldsymbol{r}_{ij}}\left(\sum_{\boldsymbol{n}}\frac{1}{\left|\boldsymbol{r}_{ij}+\boldsymbol{d}-\boldsymbol{n}L\right|}\right)\\
= & \underset{n}{\sum}\left[-\frac{1}{\left|\boldsymbol{r}_{ij}+\boldsymbol{d}-\boldsymbol{n}L\right|^{2}}\textrm{erfc}\left(\frac{\left|\boldsymbol{r}_{ij}+\boldsymbol{d}-\boldsymbol{n}L\right|}{\sqrt{2}\sigma}\right)\right.\\
 & \left.+\frac{1}{\left|\boldsymbol{r}_{ij}+\boldsymbol{d}-\boldsymbol{n}L\right|}\sqrt{\frac{2}{\pi}}\frac{1}{\sigma}\cdot e^{-\left(\frac{\left|\boldsymbol{r}_{ij}+\boldsymbol{d}-\boldsymbol{n}L\right|}{\sqrt{2}\sigma}\right)^{2}}\right]\\
 & +\frac{4\pi}{\Omega}\sum_{\boldsymbol{k}\neq0}\frac{e^{-\sigma^{2}k^{2}/2}}{k^{2}}\boldsymbol{k}\textrm{sin}\left(\boldsymbol{k}\cdot\left(\boldsymbol{r}_{ij}+\boldsymbol{d}\right)\right),
\end{align*}
which leads to 

\[
E_{\textrm{CD}}=\stackrel[i,j\alpha]{N}{\sum}q_{i}Q_{i,j\alpha}\left(\boldsymbol{u}_{j}\right)_{\alpha},
\]
where 

\begin{align}
 & Q_{i,j\alpha}\nonumber \\
= & -\frac{1}{4\pi\varepsilon_{0}}\left\{ \sum_{n}\left[\frac{\left(\boldsymbol{r}_{ij}+\boldsymbol{d}-\boldsymbol{n}L\right)_{\alpha}}{\left|\boldsymbol{r}_{ij}+\boldsymbol{d}-\boldsymbol{n}L\right|^{3}}\textrm{erfc}\left(\frac{\left|\boldsymbol{r}_{ij}+\boldsymbol{d}-\boldsymbol{n}L\right|}{\sqrt{2}\sigma}\right)\right.\right.\nonumber \\
 & \left.-\frac{\left(\boldsymbol{r}_{ij}+\boldsymbol{d}-\boldsymbol{n}L\right)_{\alpha}}{\left|\boldsymbol{r}_{ij}+\boldsymbol{d}-\boldsymbol{n}L\right|^{2}}\sqrt{\frac{2}{\pi}}\frac{1}{\sigma}\cdot e^{-\left(\frac{\left|\boldsymbol{r}_{ij}+\boldsymbol{d}-\boldsymbol{n}L\right|}{\sqrt{2}\sigma}\right)^{2}}\right]\nonumber \\
 & \left.-\frac{4\pi}{\Omega}\sum_{\boldsymbol{k}\neq0}\frac{e^{-\sigma^{2}k^{2}/2}}{k^{2}}\textrm{sin}\left[\boldsymbol{k}\cdot\left(\boldsymbol{r}_{ij}+\boldsymbol{d}\right)\right]k_{\alpha}\right\} .\label{eq:chg-dipQija}
\end{align}

\subsection{Ion displacements \label{subsec:small-shifts}}

Since dipoles arise when charges are displaced from its original positions,
one may wonder if it is reasonable to replace dipole-dipole interaction
with pure charge-charge interaction in the investigation of perovskites.
In this way, the Coulomb energy can be calculated without resorting
to dipoles. However, one possible disadvantage will be the recalculation
of the Ewald matrix each time when charges are displaced. An alternative
way is to implement the P$^{3}$M or the PME method in MC programs
where particle positions need not be fixed. However, this is likely
still too slow since every time one charge is changed, a new computation
over the whole system is needed, which implies the need to design
new and more efficient algorithms.  

One possible simplification is to deal with small charge displacements.
Given the charge-charge interaction matrix, if the charge $q_{i}$
is displaced by $\boldsymbol{s}_{i}$ to $\boldsymbol{r}_{i}-\boldsymbol{n}L+\boldsymbol{s}_{i}$,
the new interaction matrix can be obtained by expanding $Q_{ij}$
to the second order of $\boldsymbol{s}_{i}$, i.e., 
\begin{align}
 & Q_{ij}\left(\left\{ \boldsymbol{r}_{i}+\boldsymbol{s}_{i}\right\} \right)\nonumber \\
= & Q_{ij}\left(\left\{ \boldsymbol{r}_{i}\right\} \right)+\frac{1}{2}\frac{1}{4\pi\varepsilon_{0}}\left(-\frac{2\pi}{\Omega}\right)\nonumber \\
 & \times\sum_{\boldsymbol{k}\neq0}\frac{e^{-\sigma^{2}k^{2}/2}}{k^{2}}\cos\left[\boldsymbol{k}\cdot\left(\boldsymbol{r}_{i}-\boldsymbol{r}_{j}\right)\right]\left[\boldsymbol{k}\cdot\left(\boldsymbol{s}_{i}-\boldsymbol{s}_{j}\right)\right]^{2}\nonumber \\
= & Q_{ij}\left(\left\{ \boldsymbol{r}_{i}\right\} \right)-\frac{1}{2}\sum_{\alpha\beta}\left(\boldsymbol{s}_{i}-\boldsymbol{s}_{j}\right)_{\alpha}\left(\boldsymbol{s}_{i}-\boldsymbol{s}_{j}\right)_{\beta}G_{ij\alpha\beta}\label{eq:shifted-charges}
\end{align}
where
\begin{align}
G_{ij\alpha\beta}= & \left(\frac{1}{2\varepsilon_{0}\Omega}\right)\sum_{\boldsymbol{k}\neq0}\frac{e^{-\sigma^{2}k^{2}/2}}{k^{2}}\cos\left[\boldsymbol{k}\cdot\left(\boldsymbol{r}_{i}-\boldsymbol{r}_{j}\right)\right]k_{\alpha}k_{\beta}.\label{eq:Gijab}
\end{align}
We note that the expression in Eq. (\ref{eq:shifted-charges}) cannot
be reduced to the dipole-dipole interaction despite their apparent
similarity.  

\section{Implementation\label{sec:Implementation}}

For calculation involving Ewald summation, there are two common scenarios:
(i) Charges and dipoles change frequently and every calculation needs
to take into account all the changes. For instance, in a MD simulation
all charges and dipoles can change at every time step; and (ii) Only
one or just a few charges and dipoles are changed. For instance, in
a MC simulation, during one MC sweep the dipoles or charges can be
changed individually one by one. For scenario (i), the expressions
in Sec. \ref{sec:Further-simplification} can be useful. For scenario
(ii), an interaction matrix can be used to speedup the calculation. 

In Secs. \ref{sec:charge-charge} and \ref{sec:Other-interaction-matrices},
we have derived the interaction matrices for charge-charge, dipole-dipole,
charge-dipole, displaced charge interactions, which are given in Eq.
(\ref{eq:charge-charge_matrix}), Eq. (\ref{eq:dipole-dipole-matrix}),
Eq. (\ref{eq:chg-dipQija}), and Eq. (\ref{eq:shifted-charges}),
respectively. These matrices are numerically implemented using Python.
For computationally intensive parts, we have used C++ to accelerate
the calculation \citep{pybind11}. The output of the matrices are
stored in \texttt{netcdf} format for cross-platform deployment and
easy use for different type of simulations (e.g., MC and MD). The
source code can be found on \texttt{GitLab} \citep{software-package}.

\begin{table}[h]
\begin{centering}
\begin{tabular}{|c|c|c|}
\hline 
Symmetry \footnote{The symmetry notations here specifie the dipole configurations, which
can be found in Ref. {[}\onlinecite{Zhong1995}{]}.} & Ref. {[}\onlinecite{Nishimatsu2010}{]} \footnote{In unit of $Z^{\ast}/4\pi\varepsilon_{\infty}a_{0}^{3}$, where $Z^{\ast}$
is the effective charge, $a_{0}$ is the lattice constant, and $\varepsilon_{\infty}$
is the permittivity.} & Ewald\tabularnewline
\hline 
\hline 
$\Gamma$ & $-\frac{2}{3}\pi$ & -2.094\tabularnewline
\hline 
X$_{1}$ & 4.844 & 4.844\tabularnewline
\hline 
X$_{5}$ & -2.422 & -2.422\tabularnewline
\hline 
M$_{3}$ & -2.677 & -2.677\tabularnewline
\hline 
M$_{5}$ & 1.338 & 1.338\tabularnewline
\hline 
R$_{25}$ & 0 & -6.348$\times10^{-10}$\tabularnewline
\hline 
$\Sigma_{\textrm{Lo}}$ & 2.932 & 2.932\tabularnewline
\hline 
\end{tabular}
\par\end{centering}
\caption{Dipole interaction for selected high-symmetry dipole configurations.
\label{tab:Dipole-interaction-for}}
\end{table}
 To verify the numerical implementation, we performed a series of
tests \citep{software-package} : (i) For charge-charge interaction
matrix, we have used $2\times2\times2$ supercell and calculated the
Madelung constants of NaCl ($M=1.7476$) and CsCl ($M=1.7627$), obtaining
correct results \citep{Johnson1961}; (ii) We have used $4\times4\times4$
supercell to calculate the dipole-dipole interaction energy (analogy
to the Madelung constant) of selected high-symmetry dipole configurations
and compared to available results \citep{Nishimatsu2010} (see Tab.
\ref{tab:Dipole-interaction-for}); (iii) For charge-dipole interaction
matrix, we have constructed a few charge and dipole configurations,
calculated the charge-dipole electrostatic energy using the interaction
matrix, and compared to the results obtained by direct summation in
real-space \citep{real-space}. These results can be obtained by running
the Python programs found in the ``test'' directory of our \texttt{PyEwald}
program \citep{software-package}.

\section{Discussion\label{sec:Discussion}}

Having shown the derivation of main results, we proceed to discuss
the applicability of the interaction matrix to general Bravais lattices,
the summation in real space, and provide alternative expressions for
the total energy and more details important for obtaining results
in previous sections.

\subsection{General Bravais lattice\label{sec:General-Bravai-lattice}}

In previous sections, simple cubic lattice is conveniently assumed
for the Ewald summation. Here let us check if our results can be easily
extended to more general Bravais lattice. In the following, we focus
on the charge-charge interaction and use it as an example.

The key step here is to check whether Eq. (\ref{eq:sum-1/r}) is still
valid for general Bravais lattice. To this end, we show that the Poisson-Jacobi
relation is also true for general Bravais lattice. That is, we would
like to show that the equation 

\begin{align}
\sum_{L}e^{-\left|\boldsymbol{r}+\boldsymbol{L}\right|^{2}t^{2}}= & \frac{1}{V}\left(\frac{\pi}{t^{2}}\right)^{3/2}\sum_{k}e^{i\boldsymbol{k}\cdot\boldsymbol{r}}\exp\left(-\frac{k^{2}}{4t^{2}}\right),\label{eq:P-J_relation}
\end{align}
is still valid for a general Bravais lattice. With the previous equation
and the the fact that
\begin{align}
\frac{1}{\left|\boldsymbol{r}+\boldsymbol{L}_{\boldsymbol{n}}\right|}= & \frac{2}{\sqrt{\pi}}\int_{0}^{\infty}dt\exp\left(-\left|\boldsymbol{r}+\boldsymbol{L}_{\boldsymbol{n}}\right|^{2}t^{2}\right)\nonumber \\
= & \frac{2}{\sqrt{\pi}}\int_{0}^{\gamma}dt\exp\left(-\left|\boldsymbol{r}+\boldsymbol{L}_{\boldsymbol{n}}\right|^{2}t^{2}\right)\nonumber \\
 & +\frac{2}{\sqrt{\pi}}\int_{\gamma}^{\infty}dt\exp\left(-\left|\boldsymbol{r}+\boldsymbol{L}_{\boldsymbol{n}}\right|^{2}t^{2}\right)\label{eq:integral-transformation}
\end{align}
the Ewald summation formula can be derived \citealp{Ziman,Mazars2011}.
As a matter of fact, these two equations are the basis for an alternative
derivation of the Ewald summation adopted by some authors \citep{Ziman}.
Therefore, if the Eq. (\ref{eq:P-J_relation}) is true for general
Bravais lattice, then the expressions of Ewald summation are also
true for general Bravais lattice.

In the following, we outline the proof using 
\begin{align*}
f\left(\boldsymbol{r},t\right) & =\sum_{\boldsymbol{L}_{n}}e^{-\left|\boldsymbol{r}+\boldsymbol{L}_{n}\right|^{2}t^{2}},
\end{align*}
where $\boldsymbol{L}_{\boldsymbol{n}}=\boldsymbol{a}_{1}n_{1}+\boldsymbol{a}_{2}n_{2}+\boldsymbol{a}_{3}n_{3}$
and $\boldsymbol{a}_{1}$, $\boldsymbol{a}_{2}$ and $\boldsymbol{a}_{3}$
are the Bravais vectors of a general lattice and $n_{1,2,3}$ are
integers. It is easy to see that, $f\left(\boldsymbol{r}+\boldsymbol{a}_{1}n_{1}+\boldsymbol{a}_{2}n_{2}+\boldsymbol{a}_{3}n_{3},t\right)=f\left(\boldsymbol{r},t\right)$
is true for arbitrary $n_{1,2,3}$, which shows that $f\left(\boldsymbol{r},t\right)$
is a periodic function that can be expanded into a Fourier series
\begin{align*}
f\left(\boldsymbol{r},t\right)= & \sum_{m}c_{m}\left(t\right)\exp\left(i\boldsymbol{G}_{m}\cdot\boldsymbol{r}\right),
\end{align*}
with the requirement that $\boldsymbol{G}_{m}\cdot(\boldsymbol{a}_{1}n_{1}+\boldsymbol{a}_{2}n_{2}+\boldsymbol{a}_{3}n_{3})=1$
for arbitrary $n_{1,2,3}$. This condition essentially requires that
\begin{align*}
\boldsymbol{G}_{m}= & \boldsymbol{b}_{1}m_{1}+\boldsymbol{b}_{2}m_{2}+\boldsymbol{b}_{3}m_{3},
\end{align*}
 $\boldsymbol{b}_{1}$, $\boldsymbol{b}_{2}$ and $\boldsymbol{b}_{3}$
are the primitive vectors reciprocal to $\boldsymbol{a}_{1}$, $\boldsymbol{a}_{2}$
and $\boldsymbol{a}_{3}$ \citep{Ziman}. The coefficients of the
Fourier series are 
\begin{align*}
c_{m}\left(t\right)= & \frac{1}{V}\sum_{\boldsymbol{L_{n}}}\int d\boldsymbol{r}e^{-\left|\boldsymbol{r}+\boldsymbol{L_{n}}\right|^{2}t^{2}}\exp\left(-i\boldsymbol{G}_{m}\cdot\boldsymbol{r}\right)\\
= & \frac{1}{\Omega}\int d\boldsymbol{r}e^{-\boldsymbol{r}^{2}t^{2}}\exp\left(-i\boldsymbol{G}_{m}\cdot\boldsymbol{r}\right)\\
= & \frac{1}{\Omega}\left(\frac{\pi}{t^{2}}\right)^{3/2}\exp\left(-\frac{1}{4t^{2}}\left|\boldsymbol{G}_{m}\right|^{2}\right),
\end{align*}
where $\Omega$ is the volume spanned by $\boldsymbol{a}_{1}$, $\boldsymbol{a}_{2}$
and $\boldsymbol{a}_{3}$.

Therefore
\begin{align}
 & \sum_{\boldsymbol{L}_{n}}e^{-\left|\boldsymbol{r}+\boldsymbol{L}_{n}\right|^{2}t^{2}}\nonumber \\
= & \frac{1}{\Omega}\left(\frac{\pi}{t^{2}}\right)^{3/2}\sum_{m}\exp\left(i\boldsymbol{G}_{m}\cdot\boldsymbol{r}\right)\exp\left(-\frac{1}{4t^{2}}\left|\boldsymbol{G}_{m}\right|^{2}\right),\label{eq:exp-sum}
\end{align}
which is the same as Eq. (\ref{eq:P-J_relation}) with $\boldsymbol{G}_{m}$
(made of $\boldsymbol{b}_{1,2,3}$) being the reciprocal lattice of
a given general Bravais lattice. Combining the above expression and
Eq. (\ref{eq:integral-transformation}), we can finally have 
\begin{align*}
 & \sum_{n}\frac{1}{\left|\boldsymbol{r}+\boldsymbol{L}_{n}\right|}\\
= & \frac{4\pi}{\Omega}\sum_{m}\frac{\exp\left(-\left|\boldsymbol{G}_{m}\right|^{2}/4\gamma^{2}\right)}{\left|\boldsymbol{G}_{m}\right|^{2}}\cos\left(\boldsymbol{G}_{m}\cdot\boldsymbol{r}\right)\\
 & +\sum_{n}\frac{\textrm{erfc}\left(\left|\boldsymbol{r}+\boldsymbol{L}_{\boldsymbol{n}}\right|\gamma\right)}{\left|\boldsymbol{r}+\boldsymbol{L}_{\boldsymbol{n}}\right|},
\end{align*}
which is the the Ewald summation for general Bravais lattices.

\subsection{Summation in real space\label{subsec:Summation-in-real}}

It can be shown that the real-space summation for all the interaction
matrices can be ignored with proper choice of $\sigma$. Rather than
providing exact proofs, we demonstrate why they can be ignored, which
is further verified numerical simulation.

For the charge-charge interaction, we first check the second term
of Eq. (\ref{eq:charge-charge_matrix}). The largest possible value
exists when $\boldsymbol{r}_{i}=\boldsymbol{r}_{j}$ and $\boldsymbol{n}=(1,0,0)$
or equivalent terms: 
\begin{align*}
A= & \frac{1}{L}\textrm{erfc}\left(\frac{L}{\sqrt{2}\sigma}\right)=\frac{1}{L}\textrm{erfc}\left(\alpha L\right).
\end{align*}
If we choose $\alpha\approx\sqrt{-\ln\delta}=5.26$ with $\delta=10^{-12}$
and $L=1$ (for the smallest $1\times1\times1$ supercell), then $A=1.02\times10^{-13}$.
This value is comparable to the round error in numerical calculation
and decreases with $L$. Using the same argument, it can be shown
that the first term can also be safely ignored. Such practice has
been numerically verified. 

For the dipole-dipole interaction, the real-space sum is more complex
as shown in Eq. (\ref{eq:supercell-dip-dip-1}). To estimate how large
this term is, we calculate its upper bound
\begin{align*}
S\left(x\right)= & \frac{1}{\sigma^{3}x^{3}}\sqrt{\sum_{\alpha,\beta}\left[-\delta_{\alpha\beta}B\left(x\right)+\frac{3x_{\alpha}x_{\beta}}{x^{2}}C\left(x\right)\right]^{2}}\\
= & \frac{\sqrt{3}}{\sigma^{3}x^{3}}\sqrt{\left[B\left(x\right)-C\left(x\right)\right]^{2}+2C^{2}\left(x\right)}
\end{align*}
Since $S\left(x\right)$ is a monotonically decreasing function, we
can check $\left|\boldsymbol{r}\right|=1$, $\boldsymbol{n}=\left(0,0,0\right)$
( $x=\frac{\left|\boldsymbol{r}-\boldsymbol{n}L\right|}{\sqrt{2}\sigma}$),
which gives $S=9.62\times10^{-10}$ when $\alpha=1/\sqrt{2}\sigma=5.26$.
Therefore it can also be ignored in numerical calculations.

For the charge-dipole interaction, we can again estimate the real-space
summation terms in Eq. (\ref{eq:chg-dipQija}) by calculating the
largest terms $\left(\left|\boldsymbol{d}\right|=0.5\right)$. The
value of the two terms are given by

\begin{align*}
A= & \frac{1}{d^{2}}\textrm{erfc}\left(\frac{d}{\sqrt{2}\sigma}\right)=4.22\times10^{-13},\\
B= & \frac{1}{d}\sqrt{\frac{2}{\pi}}\frac{1}{\sigma}\exp\left(-\frac{d^{2}}{2\sigma^{2}}\right)=2.37\times10^{-11}.
\end{align*}
with $\alpha=1/\sqrt{2}\sigma=2\sqrt{-\ln\left(\delta\right)}=10.5$.
Here the value of $\alpha$ is doubled to compensate the reduced distance
(from 1 to 0.5).

\subsection{Other useful expressions\label{sec:Further-simplification}}

In Secs \ref{sec:charge-charge}, \ref{sec:Dipole-dipole-interaction},
and \ref{sec:Charge-dipole-interaction}, we have obtained various
interaction matrices that are suitable for simulations when charges
and dipoles are changed one by one. There are situations when charges
and dipoles are changed by large clusters, where alternative expressions
are necessary to further increase the computational efficiency.

The long-range part of the charge-charge interaction in Eq. (\ref{eq:E_L-Ewald})
can be converted to, 
\begin{align*}
 & E^{L}\\
= & \frac{1}{2}\frac{1}{4\pi\varepsilon_{0}\Omega}\sum_{\boldsymbol{k}\neq0}\frac{4\pi e^{-\sigma^{2}k^{2}/2}}{k^{2}}\left|U^{2}\left(\boldsymbol{k}\right)\right|\\
 & -\frac{1}{4\pi\varepsilon_{0}}\frac{1}{\sqrt{2\pi}\sigma}\sum_{i=1}^{N}q_{i}^{2}
\end{align*}
where
\begin{align}
U\left(\boldsymbol{k}\right)= & \sum_{i}q_{i}\exp\left(i\boldsymbol{k}\cdot\boldsymbol{r}_{i}\right).\label{eq:U_k}
\end{align}
 As we have shown in Sec. \ref{subsec:Summation-in-real}, $E^{S}$
can be ignored with proper choice of $\sigma$. Therefore the above
expression is the also the total Coulomb energy due to charge-charge
interaction.

The dipole energy can be converted to:
\begin{align*}
E_{\textrm{dip-dip}}= & \frac{1}{2}\frac{1}{4\pi\varepsilon_{0}}\left(\sqrt{\frac{2}{\pi}}\frac{1}{3\sigma^{3}}\right)\sum_{i\in\textrm{supercell}}\left|\boldsymbol{u}_{i}\right|^{2}\\
 & -\frac{1}{2\Omega\varepsilon_{0}}\sum_{\boldsymbol{k}\neq0}\frac{e^{-\sigma^{2}k^{2}/2}}{k^{2}}\left|U\left(\boldsymbol{k}\right)\right|^{2},
\end{align*}
where 
\begin{align*}
U\left(\boldsymbol{k}\right)= & \sum_{i}\left(\boldsymbol{k}\cdot\boldsymbol{u}_{i}\right)\exp\left(i\boldsymbol{k}\cdot\boldsymbol{r}_{i}\right),
\end{align*}

Similarly, the alternative expression for charge-dipole interaction
is 
\begin{align*}
E_{\textrm{CD}}^{L}= & -\frac{1}{\Omega\varepsilon_{0}}\sum_{\boldsymbol{k}\neq0}\frac{e^{-\sigma^{2}k^{2}/2}}{k^{2}}\Im U\left(\boldsymbol{k}\cdot\boldsymbol{r}_{i}+\boldsymbol{k}\cdot\boldsymbol{d}\right)V\left(\boldsymbol{k}\cdot\boldsymbol{r}_{j}\right),
\end{align*}
where 
\begin{align*}
U\left(\boldsymbol{k}\right)= & \sum_{i}q_{i}\exp\left[i\left(\boldsymbol{k}\cdot\boldsymbol{r}_{i}+\boldsymbol{k}\cdot\boldsymbol{d}\right)\right],\\
V\left(\boldsymbol{k}\right)= & \sum_{j}\left(\boldsymbol{k}\cdot\boldsymbol{u}_{j}\right)\exp\left(i\boldsymbol{k}\cdot\boldsymbol{r}_{j}\right),
\end{align*}
and $\Im$ takes the imaginary part of a complex number. 

\subsection{Derivation of $a_{\boldsymbol{k}}$\label{subsec:Derivation-of-alpha-k}}

Obtaining the Fourier coefficients of a periodic function as shown
in Eq. (\ref{eq:f-r}) is important. The coefficient is given by

\begin{align*}
a_{\boldsymbol{k}}= & \frac{1}{V}\int_{V}d\boldsymbol{r}f\left(r\right)\exp\left(-i\boldsymbol{k}\cdot\boldsymbol{r}\right)\\
= & \frac{1}{V}\sum_{\boldsymbol{n}}\int_{V}d\boldsymbol{r}\frac{1}{\left|\boldsymbol{r}-\boldsymbol{n}L\right|}\textrm{erf}\left(\frac{\left|\boldsymbol{r}-\boldsymbol{n}L\right|}{\sqrt{2}\sigma}\right)\exp\left(-i\boldsymbol{k}\cdot\boldsymbol{r}\right).
\end{align*}
Given the periodicity of the function inside the above integral and
noting $\boldsymbol{k}$ are the reciprocal lattice points, we have
\begin{align}
a_{\boldsymbol{k}}= & \frac{2\pi}{\Omega}\int_{0}^{\infty}drr\textrm{erf}\left(\frac{r}{\sqrt{2}\sigma}\right)\int_{0}^{\pi}\sin\theta d\theta\exp\left(-ikr\cos\theta\right),\label{eq:spherical-condition-used}
\end{align}
where spherical coordinate is used and the azimuth axis is set to
be along $\boldsymbol{k}$. The above expression can be further converted
to 
\begin{align*}
a_{\boldsymbol{k}}= & \frac{4\pi}{\Omega}\frac{1}{k^{2}}\int_{0}^{\infty}dx\sin x\textrm{erf}\left(\frac{x}{\sqrt{2}k\sigma}\right)\\
= & \frac{4\pi}{\Omega}\frac{1}{k^{2}}\frac{2}{\sqrt{\pi}}\int_{0}^{\alpha}dy\int_{0}^{\infty}dx\sin xx\exp\left(-x^{2}y^{2}\right)
\end{align*}
where the definition of $\textrm{erf}\left(x\right)$ has been used
and $\ensuremath{\alpha=1/\sqrt{2}k\sigma}$. Integrating over $x$first
we have
\begin{align*}
a_{\boldsymbol{k}}= & \frac{2\pi}{\Omega}\frac{1}{k^{2}}\int_{0}^{\alpha}\frac{\exp\left(-1/4y^{2}\right)}{y^{3}}dy\\
= & \frac{4\pi}{\Omega k^{2}}\exp\left(-\frac{k^{2}\sigma^{2}}{2}\right),
\end{align*}
which is the result we have used in previous sections. 

In the above expression, we note that $\boldsymbol{k}=0$ is a special
case \citep{Jackson1999}. From Eq. (\ref{eq:spherical-condition-used}),
we have
\begin{align*}
a_{0}= & \frac{4\pi}{\Omega}\int_{0}^{\infty}drr\left.\frac{\sin\left(kr\right)}{kr}\right|_{k=0}\textrm{erf}\left(\frac{r}{\sqrt{2}\sigma}\right)\\
= & \frac{8\pi\sigma^{2}}{\Omega}\int_{0}^{\infty}dxx\textrm{erf}\left(x\right)\\
\equiv & \frac{8\pi\sigma^{2}}{\Omega}X_{0}
\end{align*}
where $X_{0}=\int_{0}^{\infty}dxx\textrm{erf}\left(x\right)$ is infinitely
large. If this term is included, then the energy of Eq. (\ref{eq:E_L-Ewald})
will be added by one more term, i.e. $\frac{\sigma^{2}}{\Omega}X_{0}Q^{2}$,
where $Q=\sum_{i=1}^{N}q_{i}$ is the total charge in a supercell.
While this term is infinite, if we fix $\sigma$, $\Omega$, and assume
that the net charge $Q$ in the supercell is a constant (ideally $Q=0$
with charge neutrality), this term becomes a constant that can be
ignored in practical calculations concerning energy changes.

\section{Summary \label{sec:Conclusion}}

To simulate ferroelectric perovskites, the Coulomb potential energy
between charges, dipoles, and displaced ions needs to be calculated.
For the long-range Coulomb interaction, numerical computation requires
the use of the Ewald method. In this work, we have discussed this
method in detail, provided interaction matrices, and extended to two
less investigated situations: the interaction between charges and
dipoles, as well as between displaced ions. In addition, the Ewald
method for general Bravais lattice is also considered. We note that,
due to the regular distribution of charges or dipoles, the real-space
part in the Ewald sum can be ignored. Finally, open source Python
programs implementing these interaction matrices are available on
GitLab. These analytic results, as well as the computer programs,
will find their use in the simulation of ferrelectric materials.
\begin{acknowledgments}
This work is financially supported by the National Natural Science
Foundation of China, Grant No. 11574246, U1537210, and National Basic
Research Program of China, Grant No. 2015CB654903. X.C. thanks the
financial support from Academy of Finland Projects 308647 and European
Union's Horizon2020, Grant No. 760930. D.W. also thanks the support
from China Scholarship Council (CSC No. 201706285020).
\end{acknowledgments}

\section*{Data Availability}

The Python programs developed are hosted on \texttt{GitLab} \citep{software-package},
which are freely available. The accompanying documents to the program
can be found at \url{https://dwang5.github.io/PyEwaldDoc/}.


\begin{thebibliography}{99}
\bibitem{Possion_solver} The Poisson equation concerned here is $\nabla^{2}\phi=-\rho/\varepsilon$,
where $\rho$ is the charge density distribution, $\varepsilon$ is
the permittivity, and $\phi$ is the electrostatic potential to be
found. Efficient implementation to numerically obtain $\phi$ is an
indispensable part of modern software packages for density functional
theory calculations.
\bibitem{sigma} Note that the unit of $\sigma$ is length, which
is assumed to be 1 in the present work. To deal with the more general
case, in our numerical implementation (see Sec. \ref{sec:Implementation}),
$\sigma$ also depends on the lattice constant $a$ (not the length
of the supercell $L$), i.e., $\sigma=a/\left(\sqrt{-2\ln\delta}\right)$
(or $\alpha=\sqrt{-\ln\delta}$). This choice will ensure that, for
any lattice constant, three is no need to include the sum in real
space (see Sec. \ref{subsec:Summation-in-real}; otherwise, for small
$a$, this sum may be necessary).
\bibitem{real-space} In one case, we set charges (dipole) with alternating
sign (direction) along $z$. This configuration results in the interaction
energy of 4.7173, while direct real-space summation gives 4.727 (the
maximum charge-dipole distance $d=100$).
\end{thebibliography}
\end{document}